\documentclass{aa}  
\usepackage{natbib}
\usepackage{CJK}
\usepackage{graphicx}
\usepackage{txfonts}
\usepackage{lscape}
\usepackage{longtable} 
\usepackage{threeparttable}
\usepackage{booktabs}
\usepackage{float}
\usepackage{array} 
\usepackage[figuresright]{rotating}

\usepackage{multirow}
\usepackage{url}
 
\usepackage{subfigure}
\usepackage[colorlinks=true, linkcolor=blue, citecolor=blue, urlcolor=blue]{hyperref}

\usepackage{amsmath}

\def\vlsr  {\ifmmode {V_{\rm LSR}}\else {$V_{\rm LSR}$}\fi}

\begin{document}

\title{Molecule-dependent Abundance Behavior of Oxygen-bearing Complex
Organics in High-Mass Star-Forming Regions: A Uniform 50-source Survey}

\author{
Xuefang Xu\inst{1}
\thanks{Xuefang Xu and Mingwei He contributed equally to this work.}
\thanks{Corresponding authors:
\email{xuefang\_xu@cqu.edu.cn},
\email{qian.gou@cqu.edu.cn},
\email{hejiao@sztu.edu.cn}},
Mingwei He\inst{1},
Qian Gou\inst{1},
Jiao He\inst{2},
Junzhi Wang\inst{3},
Donghui Quan\inst{4},
Di Li\inst{5,6},
Laurent Pagani\inst{7},
Juan Li\inst{8,9},
Guoming Zhao\inst{10},
Chunguo Duan\inst{1},
Yang Lu\inst{11},
\and
Luyao Zou\inst{12}
}

\institute{
School of Chemistry and Chemical Engineering,
Chongqing Key Laboratory of Theoretical and Computational Chemistry,
Chongqing University, Chongqing 401331, China
\and
College of Engineering Physics, Shenzhen Technology University,
Shenzhen 518118, China
\and
Guangxi Key Laboratory for Relativistic Astrophysics,
Department of Physics, Guangxi University, Nanning 530004, China
\and
Department of Physics, Xi'an Jiaotong-Liverpool University,
111 Ren'ai Road, Suzhou 215123, China
\and
New Cornerstone Science Laboratory, Department of Astronomy,
Tsinghua University, Beijing 100084, China
\and
National Astronomical Observatories, Chinese Academy of Sciences,
Beijing 100012, China
\and
LUX, Observatoire de Paris, PSL Research University, CNRS,
Sorbonne Universités, 75014 Paris, France
\and
Shanghai Astronomical Observatory, Chinese Academy of Sciences,
80 Nandan Road, Shanghai 200030, China
\and
Key Laboratory of Radio Astronomy, Chinese Academy of Sciences, China
\and
School of Science, Jilin Institute of Chemical Technology,
Jilin, Jilin 132022, China
\and
Zhejiang Lab, Hangzhou, Zhejiang 311121, China
\and
Laboratoire de Physicochimie de l'Atmosph\`ere,
Universit\'e du Littoral C\^ote d'Opale,
59140 Dunkerque, France
}

\abstract
{We present a uniform IRAM-30\,m survey analysis of four oxygen-bearing complex organic molecules (COMs), methanol (CH$_3$OH), acetaldehyde (CH$_3$CHO), methyl formate (CH$_3$OCHO), and dimethyl ether (CH$_3$OCH$_3$), toward 50 high-mass star-forming regions (HMSFRs) associated with 6.7\,GHz methanol masers. Column densities were derived through a homogeneous rotation-diagram approach, with CH$_3$CN used as a proxy excitation-temperature reference when needed. In CH$_3$OH-normalized abundance-ratio space, CH$_3$OCHO/CH$_3$OH and CH$_3$OCH$_3$/CH$_3$OH show the strongest pairwise correlation, whereas the correlations involving CH$_3$CHO are weaker. No clear monotonic trends are found with Galactocentric distance or beam-averaged H$_2$ column density. Comparison with previous observations places the CH$_3$OCHO--CH$_3$OCH$_3$ behavior within the range of earlier abundance-ratio measurements, while CH$_3$CHO shows larger inter-study variation.  A representative warm-up chemical model is used only for qualitative comparison with the observed abundance ranges, which are most closely matched during the decline from the post-desorption abundance peaks in the model. These results provide homogeneous beam-averaged abundance-ratio constraints for common O-bearing COMs in high-mass star-forming regions and show that their source-to-source behavior is molecule-dependent rather than fully described by a single common abundance pattern.}
\titlerunning{Molecule-dependent Abundance Behavior of Oxygen-bearing Complex Organics in High-Mass Star-Forming Regions} \authorrunning{Xu et al.}

\maketitle

\section{Introduction} \label{sec:intro}
The origin and evolution of complex organic molecules (COMs) in space are central topics in modern astrochemistry and astrobiology \citep{Herbst2009}. COMs, commonly defined in astronomy as carbon-bearing molecules containing six or more atoms, trace the transformation of simple interstellar material into chemically richer gas and ice during star formation. Their formation and abundance variations are linked to ice processing, gas--grain exchange, and energetic radiation fields in the interstellar medium \citep[ISM;][]{Belloche2013}. Despite substantial observational and modeling progress, the mechanisms governing COM formation and chemical interconnections remain poorly constrained, especially in high-mass star-forming regions \citep[HMSFRs;][]{Garrod2006,Garrod2008}. Oxygen-bearing COMs are of particular interest because they connect methanol-rich ice chemistry with more complex organic chemistry \citep{Suzuki2018,Ruaud2016}.

\begin{figure*}[htbp]
\centering
\includegraphics[width=0.98\textwidth]{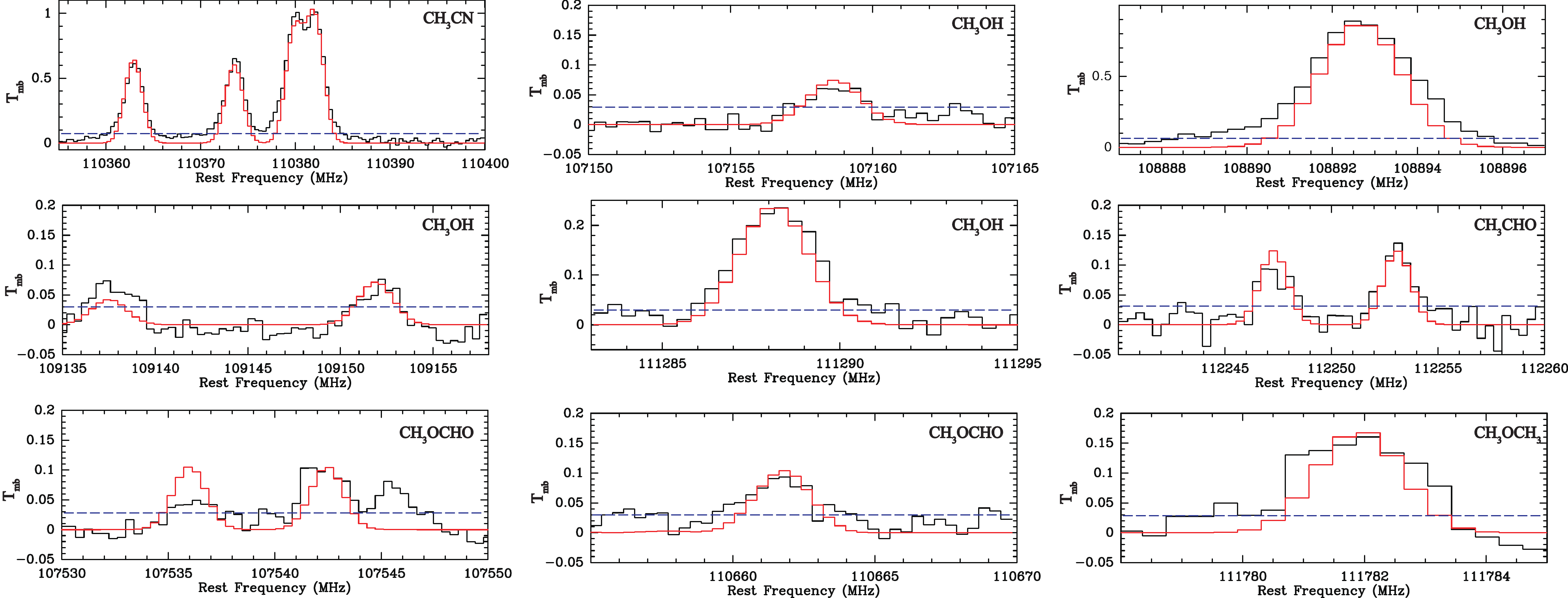}
\caption{Representative spectra toward G009.62$+$00.19 for CH$_3$CN, CH$_3$OH, CH$_3$CHO, CH$_3$OCHO, and CH$_3$OCH$_3$. The black histograms show the observed IRAM-30\,m spectra, the red curves show the corresponding LTE synthetic spectra, and the dashed horizontal lines mark the 3 rms noise level.}
\label{fig:G009.62_COMs}
\end{figure*}

Methanol (CH$_3$OH), acetaldehyde (CH$_3$CHO), methyl formate (CH$_3$OCHO), and dimethyl ether (CH$_3$OCH$_3$) are among the commonly observed O-bearing COMs in star-forming environments \citep[e.g.,][]{Bisschop2007,Vastel2014,Jorgensen2020,Biver2021}. CH$_3$OH is a key reference molecule and a major source of radicals such as CH$_3$O, CH$_3$, OH, and CH$_2$OH through photolysis, which can contribute to the formation of larger O-bearing species \citep[e.g.,][]{Garrod2006,Herbst2009,Bisschop2007,Bertin2016}. CH$_3$OCHO and CH$_3$OCH$_3$ are frequently observed together in hot cores and hot corinos, and their similar abundance behavior has often been linked to CH$_3$OH-derived chemistry, especially pathways involving the methoxy radical CH$_3$O \citep{Peeters2006,Bisschop2007,Jaber2014}. CH$_3$CHO is also chemically relevant, but its abundance behavior is not necessarily expected to follow the same pattern as CH$_3$OCHO and CH$_3$OCH$_3$.

Detailed studies of individual sources provide important insight into local processes, but source-specific conditions limit their ability to establish general abundance behavior. Uniform surveys are therefore needed to compare O-bearing COMs across larger samples. Previous studies have reported abundance relations among common O-bearing COMs in different environments \citep[e.g.,][]{Jaber2014,Coletta2020}. In particular, the IRAM-30\,m survey of 39 HMSFRs by \citet{Coletta2020} found that the CH$_3$OCHO/CH$_3$OCH$_3$ abundance ratio remains relatively stable over a broad range of source conditions, suggesting a close chemical link between these two species. However, existing constraints remain limited because sample sizes are often modest, CH$_3$CHO is not always included, and absolute abundance scales and CH$_3$OH-normalized abundance ratios are not always examined together within a uniform data set.

Motivated by these issues, we carried out a systematic study of 50 6.7\,GHz methanol-maser-associated HMSFRs based on IRAM-30\,m broadband spectral surveys. Using a homogeneous analysis of CH$_3$OH, CH$_3$CHO, CH$_3$OCHO, and CH$_3$OCH$_3$, we examine their CH$_3$OH-normalized abundance patterns across the sample. CH$_3$CN is also included as an excitation reference, not as part of the O-bearing COM family. The observations, data reduction, and line identification are presented in Sect.~\ref{sec:obser}, the results in Sect.~\ref{sec:res}, the discussion in Sect.~\ref{sec:dis}, and the main conclusions in Sect.~\ref{sec:sum}.

\begin{figure*}[!htb]
  \centering
  \includegraphics[width=0.95\textwidth]{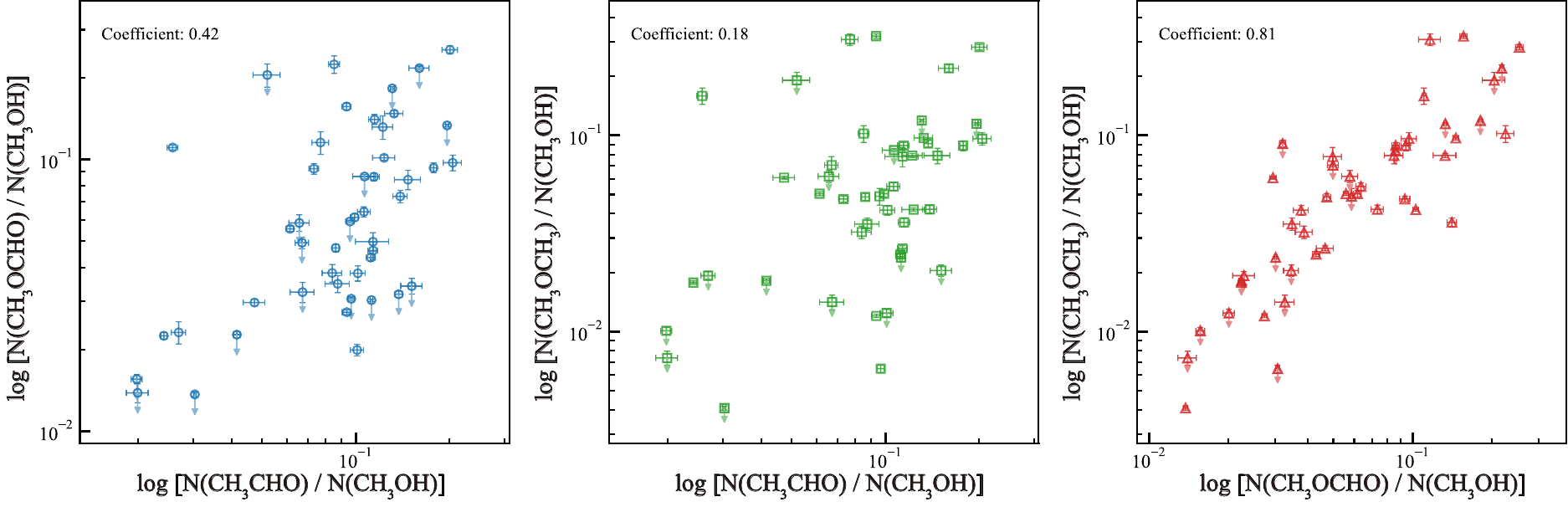}
  \caption{Pairwise correlations of CH$_3$OH-normalized abundance ratios for CH$_3$CHO, CH$_3$OCHO, and CH$_3$OCH$_3$ across the sample. Symbols with arrows indicate upper limits in the corresponding abundance ratios. Upper-limit points are shown for reference and are excluded from the correlation analysis. Pearson correlation coefficients ($\gamma$) are reported in each panel.}
  \label{fig:abundance}
\end{figure*}

\section{Observations, data reduction, and line identification}\label{sec:obser}
\subsection{Observations}

We selected 50 HMSFRs associated with 6.7\,GHz CH$_3$OH maser emission for this study. The sources are drawn from massive star-forming regions with trigonometric parallax measurements compiled by~\citet{Reid2014,Reid2019}, for which the parallaxes were measured using Class~II CH$_3$OH masers or H$_2$O masers. Following the source-selection criteria of~\citet{Liy2022}, we further required HC$_3$N $J=12$--11 emission stronger than 1\,K in main-beam brightness temperature. The 6.7\,GHz Class~II CH$_3$OH maser transition is a well-established signpost of high-mass star formation \citep[e.g.,][]{Menten1991,Minier2003,Breen2013}, while the HC$_3$N criterion ensures sufficiently bright dense-gas emission for a uniform molecular abundance analysis.

Observations were conducted with the IRAM-30\,m telescope at Pico Veleta, Spain, over three sessions: June 2016, October 2016, and August 2017. The data used in this analysis were acquired using the E0 (3\,mm) band of the Eight Mixer Receiver (EMIR), covering an 8\,GHz bandwidth at a spectral resolution of 195\,kHz (0.54\,km\,s$^{-1}$), with the Fourier Transform Spectrometer backend. A standard position-switching mode was employed with an azimuth offset of $-600''$. Further details of the observational setup are described by~\citet{Liy2022} and~\citet{Xu2025}. The source names, aliases, equatorial coordinates, and Galactocentric distances are summarized in Columns~(1)--(5) of Table~\ref{table:omd}.

\subsection{Data reduction}

At 110\,GHz (3\,mm), the IRAM-30\,m telescope provides a beam width of 22.4$''$. The main beam brightness temperature, $T_{\rm mb}$, was calculated as $T_{\rm mb} = T^{\ast}_{\rm A}{}\cdot{}F_{\rm eff} / B_{\rm eff}$, where $T^{\ast}_{\rm A}$ is the antenna temperature, $F_{\rm eff}$ is the forward efficiency, and $B_{\rm eff}$ is the main beam efficiency. At this frequency, $F_{\rm eff}$ and $B_{\rm eff}$ are 94\% and 78\%, respectively.

Data reduction was performed using the GILDAS\footnote{\url{http://www.iram.fr/IRAMFR/GILDAS}} software suite, primarily with CLASS (Continuum and Line Analysis Single-dish Software) and GREG (Grenoble Graphic). Following quality assessment, valid scans were averaged to produce a single spectrum for each source. Less than 0.5\% of the data were excluded because of baseline instabilities or faulty channels. 

Baseline subtraction was performed using first-order polynomials for most spectra. Second-order polynomials were applied to seven sources (G009.62$+$00.19, G010.47$+$00.02, G035.20$-$01.73, G059.78$+$00.06, G069.54$-$00.97, G081.75$+$00.59, and G081.87$+$00.78) to correct for mild spectral curvature. For G043.16$+$00.01 and G049.48$-$00.36, sinusoidal baselines were subtracted to mitigate low-amplitude standing-wave patterns.

\subsection{Spectral line identification}

The spectral line identification in this work builds on the 3\,mm molecular-line survey of the same 50 HMSFRs presented by \citet{Xu2025}. That study established the molecular inventory over the full 105.8--113.6\,GHz spectral coverage using the Weeds module of GILDAS \citep{Maret2011}, with spectroscopic entries from the Jet Propulsion Laboratory (JPL) and Cologne Database for Molecular Spectroscopy (CDMS) databases. In total, 27 molecular species, including 16 complex organic molecules, were identified. The present work focuses on the four target O-bearing complex organic molecules, CH$_3$OH, CH$_3$CHO, CH$_3$OCHO, and CH$_3$OCH$_3$, together with CH$_3$CN as an excitation reference.

Candidate transitions were checked against laboratory-based spectroscopic predictions, considering rest frequency, line strength, upper-state energy, expected velocity, and possible blending with other species. Synthetic LTE spectra generated with Weeds were used to verify whether the observed features could be reproduced consistently in frequency, relative intensity, linewidth, and velocity. Molecular assignments were therefore not based on single rest-frequency coincidences alone, but on the consistency of the observed spectral features and, whenever possible, multiple transitions of the same species.

For the present quantitative analysis, the Weeds synthetic spectra were used to confirm the assignments of the target O-bearing complex-organic-molecule transitions and to assess line blending. Severely blended transitions were excluded. Only unblended or minimally blended transitions with signal-to-noise ratios higher than 3 were used for the rotation-diagram analysis or for column-density estimates when a proxy temperature had to be adopted. The Weeds modeling parameters reported in Table~2 of \citet{Xu2025} were used only to guide line identification and blending assessment; the final rotation temperatures and column densities were derived independently from the rotation-diagram analysis described in Appendix~\ref{app:rd} and are listed in Table~\ref{table:oper}. Representative line-identification examples for CH$_3$CN and the four target O-bearing complex organic molecules toward G009.62$+$00.19 are shown in Fig.~\ref{fig:G009.62_COMs}. This source is retained in the quantitative analyses defined in Sect.~\ref{sec:res} and is used here as a representative line-rich case.

\section{Results}\label{sec:res}

Using the spectral line-identification procedure described in Sect.~\ref{sec:obser}, we detected CH$_3$OH, CH$_3$CHO, CH$_3$OCHO, and CH$_3$OCH$_3$ in 50, 43, 32, and 35 sources, respectively (Table~\ref{table:omd}). Thus, all four targeted O-bearing COMs have detection rates exceeding 60\% across the full sample. These detections enable a source-to-source comparison of their excitation and abundance properties.

Sgr~B2 (G000.67$-$00.03) is included in the detection statistics, but excluded from the subsequent quantitative ranges, abundance-ratio correlations, and trend analysis. Its spectra are affected by severe line crowding, foreground absorption, optical-depth effects, and multiple velocity components, preventing the same homogeneous single-component, optically thin LTE rotation-diagram treatment used for the rest of the sample. Details are given in Appendix~\ref{app:rd}. Hereafter, ``the sample'' refers to the remaining 49 sources.

Rotation temperatures ($T_{\rm rot}$) and total column densities ($N_{\rm tot}$) were obtained from rotation-diagram analysis under the optically thin LTE assumption. For molecules with a sufficient number of unblended transitions, $T_{\rm rot}$ and $N_{\rm tot}$ were derived directly from their own rotation diagrams. When such direct fits were not possible because of an insufficient number of usable transitions, limited $E_{\rm u}$ coverage, marginal detections, or non-detections, we adopted the CH$_3$CN rotation temperature measured in the same pointing as a proxy excitation temperature to estimate $N_{\rm tot}$ or upper limits.

CH$_3$CN is used as the default proxy temperature reference. It is detected in 49 sources (Table~\ref{table:omd}, Col.~6), and its $K$-ladder transitions provide a commonly used dense-gas temperature diagnostic in HMSFRs \citep[e.g.,][]{Araya2005,Hernandez2014,Gieser2019}. CH$_3$CN is not treated as part of the O-bearing COM family studied here. Instead, it provides a homogeneous excitation reference across the sample. This choice also avoids using CH$_3$OH temperatures as the default proxy for other O-bearing COMs, because CH$_3$OH is the normalization reference in the abundance-ratio analysis and may contain multiple unresolved excitation components within the single-dish beam. All proxy-temperature estimates are flagged in Table~\ref{table:oper} and should be interpreted as effective beam-averaged values rather than source-intrinsic excitation temperatures. 

\begin{figure}[!htbp]
    \centering
    \includegraphics[width=0.47\textwidth]{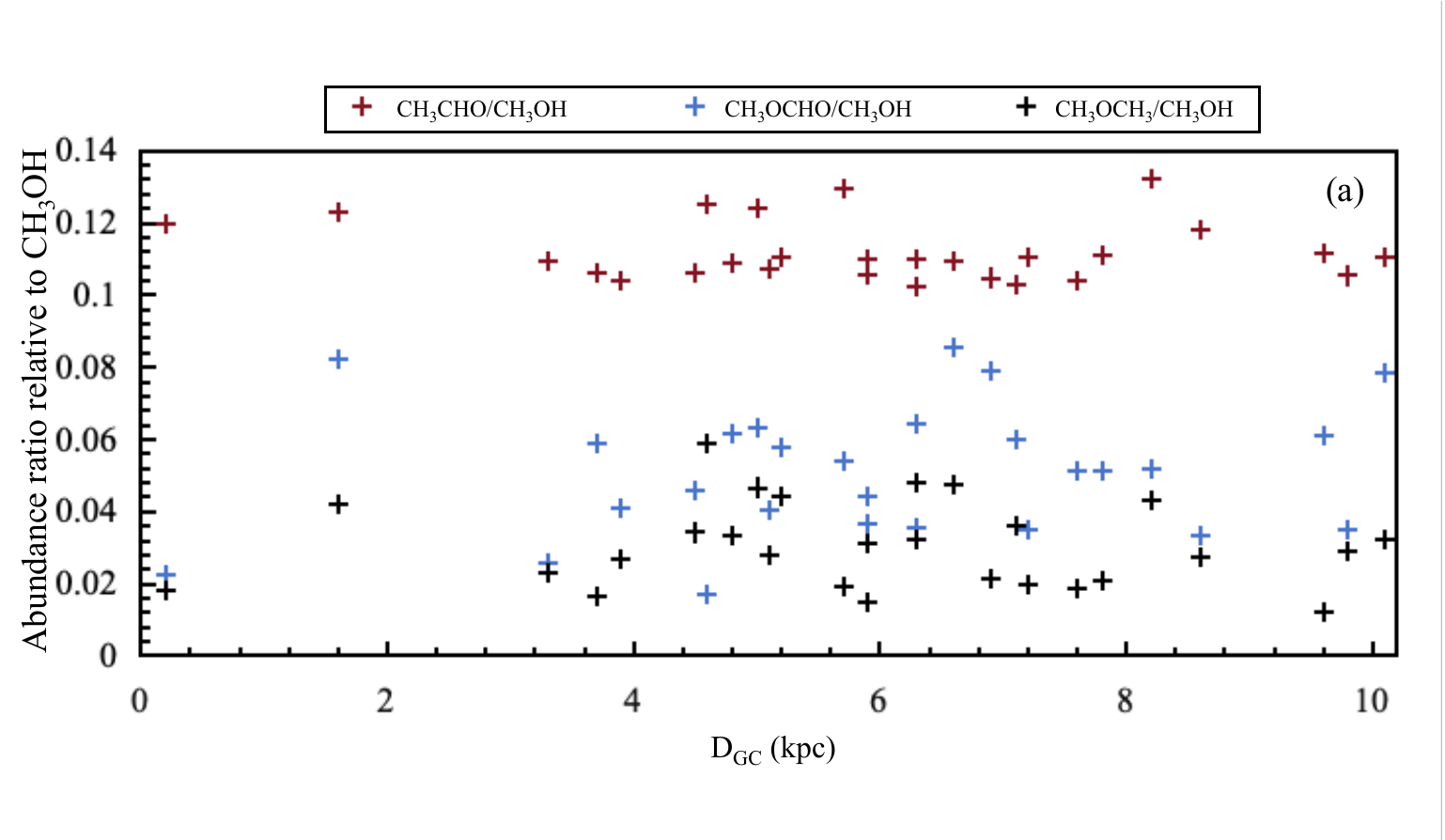}
    \includegraphics[width=0.49\textwidth]{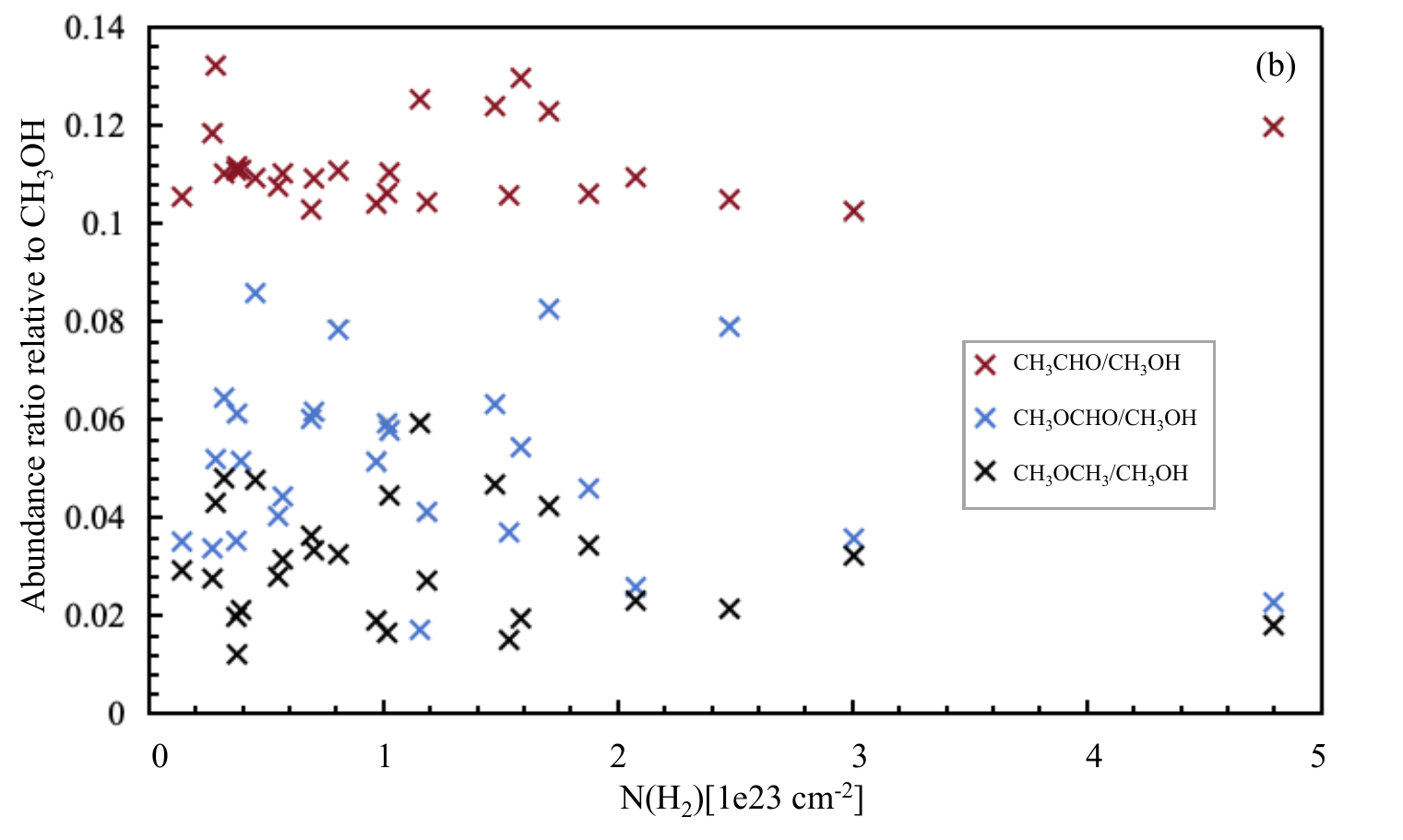}
    \caption{Abundance ratios of CH$_3$CHO, CH$_3$OCHO, and CH$_3$OCH$_3$ relative to CH$_3$OH as functions of (a) Galactocentric distance and (b) beam-averaged H$_2$ column density. No clear monotonic trends are apparent across the sampled ranges, indicating that these relative abundance ratios show no obvious dependence on the two global quantities examined here.}
    \label{fig:abundance_ratio_trends}
\end{figure}

Across the sample, CH$_3$CN gives $T_{\rm rot} \sim 30$--140\,K, whereas the O-bearing COMs span a broader range, from $\sim$30\,K to nearly 200\,K (Table~\ref{table:oper}). In some sources, O-bearing species reach rotation temperatures comparable to, or higher than, those derived from CH$_3$CN. For example, in G028.86$+$00.06, CH$_3$OH reaches $T_{\rm rot} \simeq 150$\,K, while CH$_3$CN is $\sim$70\,K. Column densities range from $\sim 10^{13}$ to a few $\times 10^{16}$\,cm$^{-2}$ across species, with CH$_3$OH being the most abundant molecule in the sample (Table~\ref{table:oper}). These values should be regarded as effective beam-averaged quantities.

To compare how the three non-methanol O-bearing COMs vary relative to a common CH$_3$OH reference, we examined CH$_3$OH-normalized abundance ratios (Fig.~\ref{fig:abundance}). In CH$_3$OH-normalized space, most abundance ratios typically span about 1--2 orders of magnitude across the sample. Here $\gamma$ denotes the Pearson correlation coefficient calculated in logarithmic abundance-ratio space using detections only; upper limits are shown in the figures for reference but are excluded from the correlation calculation. Correlations involving CH$_3$CHO are weaker, with $\gamma = 0.42$ for CH$_3$OCHO/CH$_3$OH versus CH$_3$CHO/CH$_3$OH and $\gamma = 0.18$ for CH$_3$OCH$_3$/CH$_3$OH versus CH$_3$CHO/CH$_3$OH. By contrast, CH$_3$OCHO/CH$_3$OH and CH$_3$OCH$_3$/CH$_3$OH remain well correlated, with $\gamma = 0.81$.

The $N({\rm H_2})$ values used here are adopted from~\citet{Luo2024} and provide a homogeneous beam-averaged column-density scale based on C$^{18}$O $J=1$--0 observations from the same IRAM-30\,m data set. We examined whether these relative abundance ratios vary systematically with Galactocentric distance or beam-averaged $N({\rm H_2})$ (Fig.~\ref{fig:abundance_ratio_trends}). No clear monotonic trends are found over the sampled ranges. This result only indicates that the abundance ratios show no obvious dependence on these two global quantities; other local or evolutionary factors remain unconstrained by the present analysis.

\begin{table*}[htbp]
\centering
\scriptsize
\begin{threeparttable}
\caption{Summary of literature abundance ratios used for the order-of-magnitude comparison in Fig.~\ref{fig:lit_comparison}.}
\label{tab:lit_ratio_summary}
\setlength{\tabcolsep}{3.0pt}
\renewcommand{\arraystretch}{1.20}
\begin{tabular}{p{3.3cm} p{1.8cm} p{2.3cm} p{2.4cm} p{2.4cm} p{2.5cm}}
\toprule
Reference sample
& Reference
& CH$_3$CHO/CH$_3$OH
& CH$_3$OCHO/CH$_3$OH
& CH$_3$OCH$_3$/CH$_3$OH
& CH$_3$OCHO/CH$_3$OCH$_3$ \\
\midrule

Seven high-mass young stellar objects
& \citet{Bisschop2007}
& $7.6\times10^{-6}$--$2.1\times10^{-4}$
& $0.10\pm0.03$
& $0.31\pm0.20$
& $0.19$--$0.96$ \\

G9.62$+$0.19 dense cores
& \citet{Peng2022}
& $7.1\times10^{-3}$--$1.9\times10^{-2}$
& $3.7\times10^{-2}$--$1.7\times10^{-1}$
& $6.0\times10^{-2}$--$1.5\times10^{-1}$
& $0.6$--$2.0$ \\

PEACHES low-mass protostars
& \citet{Yang2021}
& $2.3\times10^{-2}$--$9.2\times10^{-1}$
& $0.22^{+0.17}_{-0.10}$
& $0.21^{+0.15}_{-0.09}$
& $\sim 1.05$ \\

High-mass star-forming region sample
& \citet{Coletta2020}
& --
& --
& --
& $0.2$--$2.7$ \\

High-mass line-rich cores
& \citet{Li2024}
& --
& $1.5\times10^{-2}$--$1.0\times10^{-1}$
& $7.1\times10^{-3}$--$5.7\times10^{-1}$
& $0.11$--$2.68$ \\

\bottomrule
\end{tabular}

\begin{tablenotes}[flushleft]
\scriptsize
\item Notes. PEACHES denotes the Perseus ALMA Chemistry Survey. A dash indicates that the corresponding molecule was not included, or that the ratio is not available from that study. The literature values are used only as an order-of-magnitude comparison because they were obtained with different angular resolutions, excitation assumptions, source selections, and abundance definitions. For \citet{Bisschop2007}, the CH$_3$CHO/CH$_3$OH ratio is used only as a broad reference because CH$_3$CHO was treated as a colder component in that work. For \citet{Coletta2020}, only the CH$_3$OCHO/CH$_3$OCH$_3$ ratio is used; the reported mean value is $1.1\pm0.7$. For \citet{Li2024}, the median values are $3.1\times10^{-2}$, $2.5\times10^{-2}$, and $1.07$ for CH$_3$OCHO/CH$_3$OH, CH$_3$OCH$_3$/CH$_3$OH, and CH$_3$OCHO/CH$_3$OCH$_3$, respectively.
\end{tablenotes}
\end{threeparttable}
\end{table*}

\begin{figure}[htbp]
\centering
\includegraphics[width=0.35\textwidth]{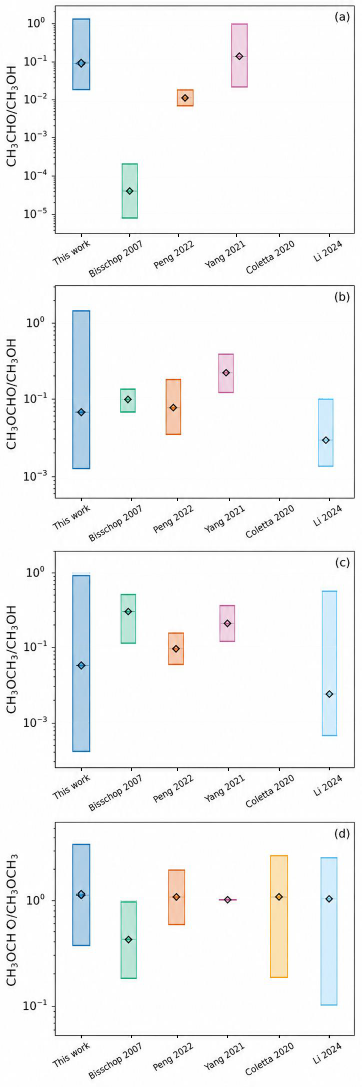}
\caption{Comparison of CH$_3$OH-normalized abundance ratios and the CH$_3$OCHO/CH$_3$OCH$_3$ ratio in this work and selected literature studies. The literature labels on the x-axis correspond to the source samples and references listed in Table~\ref{tab:lit_ratio_summary}. The panels show (a) CH$_3$CHO/CH$_3$OH, (b) CH$_3$OCHO/CH$_3$OH, (c) CH$_3$OCH$_3$/CH$_3$OH, and (d) CH$_3$OCHO/CH$_3$OCH$_3$. Colored floating bars indicate reported or recalculated ranges, while black horizontal ticks and symbols mark the median value for this work or representative literature values. Upper limits in this work are not included in the plotted ranges.}
\label{fig:lit_comparison}
\end{figure}

\section{Discussion} \label{sec:dis}
\subsection{Comparison with previous observational studies}
\label{sec:comp}

Figure~\ref{fig:lit_comparison} compares the abundance ratios derived in this work with selected literature measurements compiled in Table~\ref{tab:lit_ratio_summary}. The comparison is intended as an order-of-magnitude observational reference because the literature values were obtained with different angular resolutions, source selections, excitation treatments, and abundance definitions.

For CH$_3$OCHO and CH$_3$OCH$_3$, the ratios measured in this work overlap with the ranges reported for high-mass young stellar objects, G9.62$+$0.19 dense cores, PEACHES low-mass protostars, and high-mass line-rich cores \citep{Bisschop2007,Peng2022,Yang2021,Li2024}. The CH$_3$OCHO/CH$_3$OCH$_3$ ratio is also close to unity in these samples and in the high-mass star-forming-region sample of \citet{Coletta2020}. This is consistent with previous observational studies in which methyl formate and dimethyl ether show similar abundance behavior in chemically rich sources \citep[e.g.,][]{Brouillet2013,Jaber2014,Coletta2020,Li2024}.

The CH$_3$CHO comparison shows larger inter-study variation. The CH$_3$CHO/CH$_3$OH range in this work overlaps with the values reported for G9.62$+$0.19 and the PEACHES sources, but is higher than the range inferred for the high-mass young stellar objects of \citet{Bisschop2007}. This offset should not be interpreted directly as a chemical difference, because CH$_3$CHO was treated as a colder component in that study, whereas the present ratios are beam-averaged quantities derived within a uniform rotation-diagram analysis. Within the compiled comparison, CH$_3$CHO is therefore less consistently tied to CH$_3$OH than CH$_3$OCHO and CH$_3$OCH$_3$ are. This behavior is also compatible with the uncertain grain-surface formation efficiency of CH$_3$CHO, because quantum-chemical studies have shown that the CH$_3$ + HCO route on icy grains is sensitive to surface geometry, diffusion assumptions, and competing product channels \citep{Enrique2016,Enrique2021}.

Several additional studies are used to place the ratio comparison in a broader source- and scale-dependent context, but are not included in Fig.~\ref{fig:lit_comparison}. IRAS~16293 A/B hot-corino measurements probe O-bearing COMs on smaller spatial scales \citep{Jorgensen2018,Manigand2020}; the Fifty AU STudy of the chemistry in the disk/envelope system of Solar-like protostars (FAUST) provides CH$_3$OH-normalized O-bearing COM ratios in low-mass protostellar sources \citep{Vastel2024}; and James Webb Space Telescope (JWST)/Atacama Large Millimeter/submillimeter Array (ALMA) ice--gas comparisons constrain O-bearing COM ratios relative to CH$_3$OH in different physical phases \citep{Chen2024}. Related surveys of carbon-chain and sulfur-bearing species further show that abundance behavior can depend on molecular class and physical conditions \citep{Xie2021,Luo2019}.

\begin{figure}[htbp]
\centering
\includegraphics[width=0.48\textwidth]{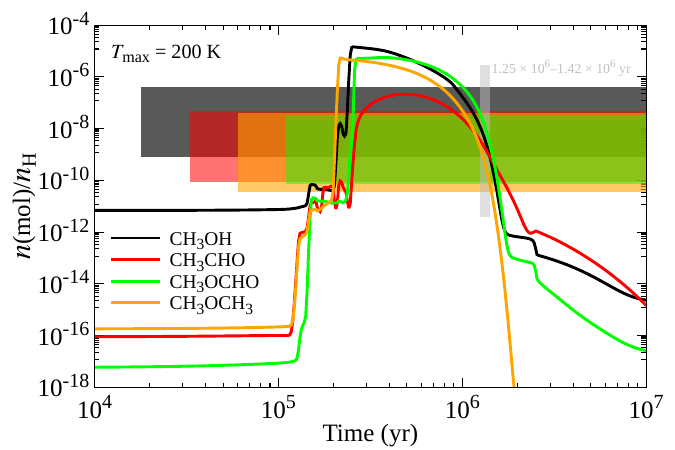}
\caption{Gas-phase abundance evolution of CH$_3$OH, CH$_3$CHO, CH$_3$OCHO, and CH$_3$OCH$_3$ in the representative warm-up model with $T_{\max}=200$\,K. The shaded rectangles indicate the observed abundance ranges derived in this work. The gray vertical band marks the time interval, $(1.25$--$1.42)\times10^{6}$\,yr, over which the model abundances overlap most closely with the observed ranges for the four species.}
\label{fig:modeling}
\end{figure}

\subsection{Qualitative comparison with warm-up chemical models}
\label{sec:modeling}

The chemical modeling was carried out with the 2024 version of \textsc{NAUTILUS} \citep{Ruaud2016,Wakelam2024}. We adopted the two-stage physical setup described by \citet{Zhao2025}, in which an initial cold phase is followed by a warm-up phase reaching $T_{\max}=200$\,K. This maximum temperature is used as a representative hot-core endpoint rather than fitted to individual sources. The resulting gas-phase abundance curves are compared with the observed beam-averaged abundance ranges in Fig.~\ref{fig:modeling}. This comparison is intended to place the observed abundance scale in the context of a representative hot-core warm-up calculation. 

In the model, the gas-phase abundances of CH$_3$OH, CH$_3$CHO, CH$_3$OCHO, and CH$_3$OCH$_3$ increase rapidly during the warm-up phase. This rise to the peak abundances is associated with efficient thermal desorption of ice-mantle material into the gas phase. The model abundances overlap most closely with the observed abundance ranges around $(1.25$--$1.42)\times10^{6}$\,yr, during their decline from the post-desorption maxima. This decline indicates that, in the model, gas-phase destruction and further chemical processing have begun to reduce the abundances after desorption.

This comparison suggests that the observed ranges are broadly consistent with a post-desorption phase of hot-core chemistry, after substantial ice-mantle desorption and during the early decline from the post-desorption abundance maxima. At this stage in the model, gas-phase processing has begun to lower the molecular abundances, but has not yet driven them to the much lower late-time values. This interpretation remains qualitative because the observations are beam-averaged, the model represents one specific physical and chemical setup, and only four species are compared. The model comparison is therefore used to contextualize the observations and is not intended to infer a unique evolutionary time for the sources or specific dominant reaction pathways.

\section{Summary} \label{sec:sum}
The main observational advance of this work is a homogeneous 50-source comparison of common O-bearing complex organic molecules (COMs). The sample consists of 6.7\,GHz methanol-maser-associated high-mass star-forming regions. We analyzed methanol (CH$_3$OH), acetaldehyde (CH$_3$CHO), methyl formate (CH$_3$OCHO), and dimethyl ether (CH$_3$OCH$_3$) using a uniform rotation-diagram approach and compared their CH$_3$OH-normalized abundance ratios across the sample. CH$_3$OCHO/CH$_3$OH and CH$_3$OCH$_3$/CH$_3$OH show the strongest pairwise correlation, whereas correlations involving CH$_3$CHO are weaker. This indicates molecule-dependent source-to-source behavior within this set of O-bearing COMs.

The CH$_3$OH-normalized ratios show no clear monotonic trends with Galactocentric distance or beam-averaged H$_2$ column density over the sampled ranges. The comparison with previous observational studies places the CH$_3$OCHO--CH$_3$OCH$_3$ behavior within the range of earlier abundance-ratio measurements, while CH$_3$CHO shows larger inter-study variation. A representative warm-up model provides qualitative context for the observed abundance ranges, which are most closely matched after substantial ice desorption in the model. These results provide homogeneous beam-averaged abundance-ratio constraints for O-bearing COM chemistry in high-mass star-forming regions, without requiring a single common abundance pattern or a unique chemical evolutionary sequence.

\section*{Data availability}
The dataset PDF and the spectral plots covering the frequency ranges
presented in this work are available through Zenodo at
\url{https://doi.org/10.5281/zenodo.19104458}.

\begin{acknowledgements}
This work is supported by the National SKA Program of China (No. 2025SKA0120100), National Natural Science Foundation of China (Grant No. W2512014, 12573065), Fundamental Research Funds for the Central Universities (Grant No. 2025CDJ-IAISYB-060), and Postdoctoral Fellows Excellence Support Program (Grant No. 2404013554893087), and the Natural Science Foundation of Top Talent of SZTU (Grant No. GDRC202402). D.L. is supported by the new Cornerstone Science Foundation.
\end{acknowledgements}

\bibliographystyle{aa} 
 \bibliography{aa60057-26} 

\begin{appendix}
\section{Rotation temperature and column density derivation}
\label{app:rd}

This appendix summarizes the procedure used to derive rotation temperatures ($T_{\rm rot}$) and beam-averaged total column densities ($N_{\rm tot}$) for CH$_3$CN and the four O-bearing complex organic molecules (COMs) considered in this work: CH$_3$OH, CH$_3$CHO, CH$_3$OCHO, and CH$_3$OCH$_3$. A table of the transitions used for each species in the source sample is included in the dataset described in the Data availability section. This table includes the molecular species, transition quantum numbers, rest frequencies, upper-state energies ($E_u$), dipole-weighted line strengths ($\mu^2S$), peak intensities ($T_{\rm mb}$), centroid velocities ($V_{\rm LSR}$), linewidths ($\Delta V$), and integrated intensities ($\int T_{\rm mb}\,dv$). We assume optically thin emission in local thermodynamic equilibrium (LTE), such that $T_{\rm rot}=T_{\rm ex}$.

For the rotation-diagram analysis, the adopted rest frequencies, Einstein $A_{\rm ul}$ coefficients, upper-state energies, degeneracies, and partition functions were taken from the same JPL and CDMS spectroscopic entries used for line identification \citep{Pickett1998,Muller2001,Muller2005}. For each molecular species, the partition function $Q(T)$ was taken from the same database as the adopted transition parameters to ensure internal consistency.

\begin{figure}[htbp]
 \centering
 \includegraphics[width=0.40\textwidth]{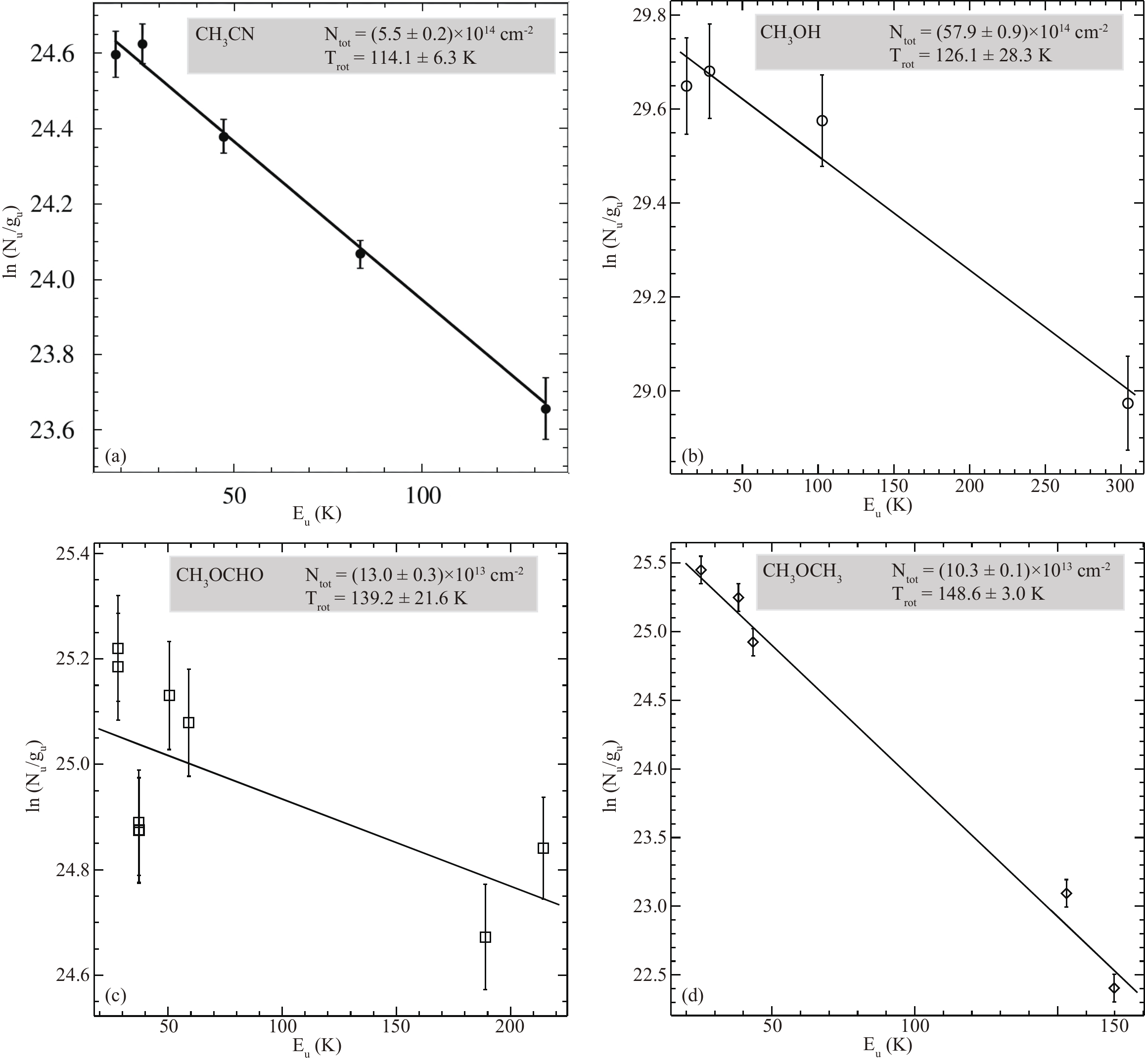}
	\caption{Rotation diagrams of detected (a) CH$_3$CN, (b) CH$_3$OH, (c) CH$_3$OCHO, and (d) CH$_3$OCH$_3$ 
		with unblended transitions in G009.62$+$00.19.}
	\label{fig:rotationaldam}           
\end{figure}

Representative rotation-diagram fits for CH$_3$CN, CH$_3$OH, CH$_3$OCHO, and CH$_3$OCH$_3$ in G009.62$+$00.19 are shown in Fig.~\ref{fig:rotationaldam}. The number of usable transitions varies from source to source depending on detection significance, blending, and the available $E_u$ coverage within the observed frequency setup. In particular, CH$_3$OH may contain multiple unresolved excitation components within the single-dish beam, so the rotation-diagram analysis adopted here should be regarded as providing an effective beam-averaged temperature and column density.

When at least three unblended or minimally blended transitions were available for a given species, we derived $T_{\rm rot}$ and $N_{\rm tot}$ from rotation diagrams under the optically thin LTE assumption. All independent rotation-diagram fits retained in Table~\ref{table:oper} are based on at least three usable transitions. For these cases, we fitted
\begin{equation}
\ln\!\left(\frac{N_u}{g_u}\right)
= -\frac{E_u}{k\,T_{\rm rot}} + \ln\!\left(\frac{N_{\rm tot}}{Q(T_{\rm rot})}\right),
\label{eq:rd}
\end{equation}
where $N_u$ is the beam-averaged column density in the upper state, $g_u$ is the statistical weight, and $k$ is the Boltzmann constant. For each transition,
\begin{equation}
N_u
= \frac{8\pi k \nu^2}{h c^3 A_{\rm ul}} \int T_{\rm mb}\, dv,
\label{eq:Nu}
\end{equation}
where $\nu$ is the rest frequency, $h$ is Planck's constant, and $c$ is the speed of light. The slope of Eq.~(\ref{eq:rd}) gives $T_{\rm rot}$, while the intercept yields $N_{\rm tot}$. Once $T_{\rm rot}$ is known, the total column density can also be written as
\begin{equation}
N_{\rm tot}
= \frac{N_u}{g_u}\, Q(T_{\rm rot})\, \exp\!\left(\frac{E_u}{kT_{\rm rot}}\right).
\label{eq:Ntot}
\end{equation}

For cases with fewer than three usable transitions, namely one or two detected transitions, $T_{\rm rot}$ was not fitted independently. Instead, $N_{\rm tot}$ was estimated by adopting a fixed excitation temperature, usually the CH$_3$CN rotation temperature measured in the same pointing. These fixed-temperature estimates are flagged in Table~\ref{table:oper}.

For non-detections, we estimated a $3\sigma$ upper limit on the integrated intensity by assuming a representative linewidth $\Delta v$, typically taken from detected lines in the same spectrum, for example from CH$_3$CN when available:
\begin{equation}
\left(\int T_{\rm mb}\, dv\right)_{3\sigma}
= 3\,{\rm rms}\,\sqrt{\Delta v\,\delta v},
\label{eq:I3sig}
\end{equation}
where ${\rm rms}$ is the per-channel noise and $\delta v$ is the channel spacing. This was converted into an upper limit on $N_u$ through Eq.~(\ref{eq:Nu}) and then into an upper limit on $N_{\rm tot}$ through Eq.~(\ref{eq:Ntot}), using the adopted excitation temperature, usually the CH$_3$CN proxy temperature when no species-specific $T_{\rm rot}$ was available.

All rotation temperatures and column densities derived from the IRAM-30\,m spectra are beam-averaged quantities. The 22$''$ beam can encompass unresolved substructure, including multiple dense cores, in many HMSFRs. Therefore, the derived values should not be interpreted as excitation temperatures or column densities of individual compact cores. This limitation affects the absolute values of $T_{\rm rot}$ and $N_{\rm tot}$ and may contribute to the source-to-source scatter.

Sgr~B2 is excluded from the quantitative ranges and regression analyses because it represents an extreme case for which a single-component, optically thin LTE rotation-diagram treatment is not physically representative. This exclusion is not based on source multiplicity alone, but on the combination of severe line crowding, foreground absorption, optical-depth effects, and multiple velocity components, which prevents the same homogeneous selection of usable transitions adopted for the rest of the sample. Other sources may also contain unresolved multiplicity within the single-dish beam, but their target transitions can still be identified and analyzed consistently in the beam-averaged framework used for the statistical comparison. The resulting $T_{\rm rot}$ and $N_{\rm tot}$ values, together with upper limits, are listed in Table~\ref{table:oper}.

\section{Source information and derived molecular properties}
\label{app:tables}

\begin{table*}[!htb]
	\caption{Information on the sources and summary of detected CH$_{3}$CN, CH$_{3}$OH, CH$_{3}$CHO, 
		CH$_{3}$OCHO, and CH$_{3}$OCH$_{3}$.}
	\label{table:omd}
	\scriptsize
	\centering
	\begin{tabular}{llcccccccc}    
		\hline
		\hline
		Source & Alias & R.A. & Dec. & D$\rm_{GC}$* & CH$_{3}$CN & CH$_{3}$OH & CH$_{3}$CHO & CH$_{3}$OCHO & CH$_{3}$OCH$_{3}$ \\
		{} & {} & (hh:mm:ss) & (dd:mm:ss) & (kpc) & {} & {} & {} & {} & {}\\ 
		\hline
		G000.67$-$00.03 & Sgr B2            & 17:47:20.00 & $-$28:22:40.0 &  0.2 & $\surd$  & $\surd$ & $\surd$  & $\surd$  & $\surd$  \\ 
		G005.88$-$00.39 &                   & 18:00:30.31 & $-$24:04:04.5 &  5.3 & $\surd$  & $\surd$ & $\times$ & $\surd$  & $\surd$  \\ 
		G009.62$+$00.19 &                   & 18:06:14.66 & $-$20:31:31.7 &  3.3 & $\surd$  & $\surd$ & $\surd$  & $\surd$  & $\surd$  \\ 
		G010.47$+$00.02 &                   & 18:08:38.23 & $-$19:51:50.3 &  1.6 & $\surd$  & $\surd$ & $\surd$  & $\surd$  & $\surd$  \\ 
		G010.62$-$00.38 & W31               & 18:10:28.55 & $-$19:55:48.6 &  3.8 & $\surd$  & $\surd$ & $\surd$  & $\surd$  & $\surd$  \\ 
		G011.49$-$01.48 &                   & 18:16:22.13 & $-$19:41:27.2 &  7.1 & $\surd$  & $\surd$ & $\times$ & $\times$ & $\times$ \\ 
		G011.91$-$00.61 &                   & 18:13:58.12 & $-$18:54:20.3 &  5.1 & $\surd$  & $\surd$ & $\surd$  & $\surd$  & $\surd$  \\ 
		G012.80$-$00.20 &                   & 18:14:14.23 & $-$17:55:40.5 &  5.5 & $\surd$  & $\surd$ & $\surd$  & $\times$ & $\surd$  \\ 
		G012.88$+$00.48 & IRAS 18089$-$1732 & 18:11:51.42 & $-$17:31:29.0 &  5.9 & $\surd$  & $\surd$ & $\surd$  & $\surd$  & $\surd$  \\ 
		G012.90$-$00.24 &                   & 18:14:34.42 & $-$17:51:51.9 &  5.9 & $\surd$  & $\surd$ & $\times$ & $\times$ & $\times$ \\ 
		G012.90$-$00.26 &                   & 18:14:39.57 & $-$17:52:00.4 &  5.9 & $\surd$  & $\surd$ & $\surd$  & $\surd$  & $\surd$  \\ 
		G014.33$-$00.64 &                   & 18:18:54.67 & $-$16:47:50.3 &  7.2 & $\surd$  & $\surd$ & $\surd$  & $\surd$  & $\surd$  \\ 
		G015.03$-$00.67 & M17               & 18:20:24.81 & $-$16:11:35.3 &  6.4 & $\surd$  & $\surd$ & $\times$ & $\times$ & $\times$ \\ 
		G016.58$-$00.05 &                   & 18:21:09.08 & $-$14:31:48.8 &  5.0 & $\surd$  & $\surd$ & $\surd$  & $\surd$  & $\surd$  \\ 
		G023.00$-$00.41 &                   & 18:34:40.20 & $-$09:00:37.0 &  4.5 & $\surd$  & $\surd$ & $\surd$  & $\surd$  & $\surd$  \\ 
		G023.44$-$00.18 &                   & 18:34:39.19 & $-$08:31:25.4 &  3.7 & $\surd$  & $\surd$ & $\surd$  & $\surd$  & $\surd$  \\ 
		G027.36$-$00.16 &                   & 18:41:51.06 & $-$05:01:43.4 &  3.9 & $\surd$  & $\surd$ & $\surd$  & $\surd$  & $\surd$  \\ 
		G028.86$+$00.06 &                   & 18:43:46.22 & $-$03:35:29.6 &  4.0 & $\surd$  & $\surd$ & $\surd$  & $\times$ & $\surd$  \\ 
		G029.95$-$00.01 & W43S              & 18:46:03.74 & $-$02:39:22.3 &  4.6 & $\surd$  & $\surd$ & $\surd$  & $\surd$  & $\surd$  \\ 
		G031.28$+$00.06 &                   & 18:48:12.39 & $-$01:26:30.7 &  5.2 & $\surd$  & $\surd$ & $\surd$  & $\surd$  & $\surd$  \\ 
		G031.58$+$00.07 & W43Main           & 18:48:41.68 & $-$01:09:59.0 &  4.9 & $\surd$  & $\surd$ & $\surd$  & $\surd$  & $\surd$  \\ 
		G032.04$+$00.05 &                   & 18:49:36.58 & $-$00:45:46.9 &  4.8 & $\surd$  & $\surd$ & $\surd$  & $\surd$  & $\surd$  \\ 
		G034.39$+$00.22 &                   & 18:53:18.77 & $+$01:24:08.8 &  7.1 & $\surd$  & $\surd$ & $\surd$  & $\surd$  & $\surd$  \\ 
		G035.02$+$00.34 &                   & 18:54:00.67 & $+$02:01:19.2 &  6.5 & $\surd$  & $\surd$ & $\surd$  & $\times$ & $\times$ \\ 
		G035.19$-$00.74 &                   & 18:58:13.05 & $+$01:40:35.7 &  6.6 & $\surd$  & $\surd$ & $\surd$  & $\surd$  & $\surd$  \\ 
		G035.20$-$01.73 &                   & 19:01:45.54 & $+$01:13:32.5 &  5.9 & $\surd$  & $\surd$ & $\surd$  & $\times$ & $\times$ \\ 
		G037.43$+$01.51 &                   & 18:54:14.35 & $+$04:41:41.7 &  6.9 & $\surd$  & $\surd$ & $\surd$  & $\surd$  & $\surd$  \\ 
		G043.16$+$00.01 & W49N              & 19:10:13.41 & $+$09:06:12.8 &  7.6 & $\surd$  & $\surd$ & $\surd$  & $\surd$  & $\surd$  \\ 
		G043.79$-$00.12 & OH 43.8$-$0.1     & 19:11:53.99 & $+$09:35:50.3 &  5.7 & $\surd$  & $\surd$ & $\surd$  & $\surd$  & $\surd$  \\ 
		G049.48$-$00.36 & W51 IRS2          & 19:23:39.82 & $+$14:31:05.0 &  6.3 & $\surd$  & $\surd$ & $\surd$  & $\surd$  & $\surd$  \\ 
		G049.48$-$00.38 & W51M              & 19:23:43.87 & $+$14:30:29.5 &  6.3 & $\surd$  & $\surd$ & $\surd$  & $\surd$  & $\surd$  \\ 
		G059.78$+$00.06 &                   & 19:43:11.25 & $+$23:44:03.3 &  7.5 & $\surd$  & $\surd$ & $\surd$  & $\times$ & $\times$ \\ 
		G069.54$-$00.97 & ON1               & 20:10:09.07 & $+$31:31:36.0 &  7.8 & $\surd$  & $\surd$ & $\surd$  & $\surd$  & $\surd$  \\ 
		G075.76$+$00.33 &                   & 20:21:41.09 & $+$37:25:29.3 &  8.2 & $\surd$  & $\surd$ & $\surd$  & $\times$ & $\times$ \\ 
		G078.12$+$03.63 & IRAS 20126+4104   & 20:14:26.07 & $+$41:13:32.7 &  8.1 & $\surd$  & $\surd$ & $\times$ & $\times$ & $\times$ \\ 
		G081.75$+$00.59 & DR21              & 20:39:01.99 & $+$42:24:59.3 &  8.2 & $\surd$  & $\surd$ & $\surd$  & $\times$ & $\times$ \\ 
		G081.87$+$00.78 & W75N              & 20:38:36.43 & $+$42:37:34.8 &  8.2 & $\surd$  & $\surd$ & $\surd$  & $\surd$  & $\surd$  \\ 
		G092.67$+$03.07 &                   & 21:09:21.73 & $+$52:22:37.1 &  8.5 & $\surd$  & $\surd$ & $\surd$  & $\times$ & $\surd$  \\ 
		G109.87$+$02.11 & Cep A             & 22:56:18.10 & $+$62:01:49.5 &  8.6 & $\surd$  & $\surd$ & $\surd$  & $\times$ & $\times$ \\ 
		G111.54$+$00.77 & NGC 7538          & 23:13:45.36 & $+$61:28:10.6 &  9.6 & $\surd$  & $\surd$ & $\surd$  & $\surd$  & $\surd$  \\ 
		G121.29$+$00.65 & L1287             & 00:36:47.35 & $+$63:29:02.2 &  8.8 & $\surd$  & $\surd$ & $\surd$  & $\surd$  & $\times$ \\ 
		G123.06$-$06.30 & NGC281            & 00:52:24.70 & $+$56:33:50.5 &  10.1& $\surd$  & $\surd$ & $\surd$  & $\surd$  & $\surd$  \\ 
		G133.94$+$01.06 & W3OH              & 02:27:03.82 & $+$61:52:25.2 &  9.8 & $\surd$  & $\surd$ & $\surd$  & $\surd$  & $\surd$  \\ 
		G168.06$+$00.82 & IRAS 05137+3919   & 05:17:13.74 & $+$39:22:19.9 &  15.9& $\times$  & $\surd$ & $\times$ & $\times$ & $\times$ \\ 
		G176.51$+$00.20 &                   & 05:37:52.14 & $+$32:00:03.9 &  9.3 & $\surd$  & $\surd$ & $\surd$  & $\times$ & $\surd$  \\ 
		G183.72$-$03.66 &                   & 05:40:24.23 & $+$23:50:54.7 &  10.0& $\surd$  & $\surd$ & $\surd$  & $\times$ & $\surd$  \\ 
		G188.94$+$00.88 & S 252             & 06:08:53.35 & $+$21:38:28.7 &  10.4& $\surd$  & $\surd$ & $\surd$  & $\times$ & $\times$ \\ 
		G192.60$-$00.04 & S 255             & 06:12:54.02 & $+$17:59:23.3 &  9.9 & $\surd$  & $\surd$ & $\surd$  & $\surd$  & $\times$ \\ 
		G209.00$-$19.38 & Orion Nebula      & 05:35:15.80 & $-$05:23:14.1 &  8.6 & $\surd$  & $\surd$ & $\surd$  & $\surd$  & $\surd$  \\ 
		G232.62$+$00.99 &                   & 07:32:09.78 & $-$16:58:12.8 &  9.4 & $\surd$  & $\surd$ & $\times$ & $\times$ & $\times$ \\ 
		\hline
	\end{tabular}
	\tablefoot{*The Galactocentric distances of the sources are adopted from~\citet{Liy2022}. $\surd$ and $\times$ represent detection and non-detection for molecules, respectively.}
\end{table*}

\begin{sidewaystable*}[htbp]
\scriptsize
\centering
\begin{threeparttable}
\caption{Rotation temperatures and total column densities of O-bearing COMs and CH$_3$CN.}
\label{table:oper}
\setlength{\tabcolsep}{4pt}
\begin{tabular}{l cc cc cc cc cc c}
\toprule
Sources 
& \multicolumn{2}{c}{CH$_3$OH} 
& \multicolumn{2}{c}{CH$_3$CHO} 
& \multicolumn{2}{c}{CH$_3$OCHO} 
& \multicolumn{2}{c}{CH$_3$OCH$_3$} 
& \multicolumn{2}{c}{CH$_3$CN} 
& $N({\rm H_2})^{a}$ \\
\cmidrule(lr){2-3}\cmidrule(lr){4-5}\cmidrule(lr){6-7}\cmidrule(lr){8-9}\cmidrule(lr){10-11}\cmidrule(lr){12-12}
 & $T_{\rm rot}$ (K) & $N_{\rm tot}$ (cm$^{-2}$) 
 & $T_{\rm rot}$ (K) & $N_{\rm tot}$ (cm$^{-2}$)
 & $T_{\rm rot}$ (K) & $N_{\rm tot}$ (cm$^{-2}$)
 & $T_{\rm rot}$ (K) & $N_{\rm tot}$ (cm$^{-2}$)
 & $T_{\rm rot}$ (K) & $N_{\rm tot}$ (cm$^{-2}$)
 & (cm$^{-2}$) \\
\midrule
G000.67$-$00.03 & $256\pm4$   & $(4.7\pm0.5)\times10^{16}$          & $256^{\ddag}$ & $(5.6\pm0.6)\times10^{15}$               & $251\pm55$  & $(1.1\pm0.4)\times10^{15}$            & $256\pm26$  & $(8.4\pm0.5)\times10^{14}$                 & $-$          & $-$                   & $(48.0\pm0.2)\times10^{22}$ \\
G005.88$-$00.39 & $143\pm10$  & $(57.4\pm0.6)\times10^{14}$         & $60^{\ddag}$  & $1.7\times10^{14\,*}$                   & $60^{\ddag}$          & $(7.8\pm0.1)\times10^{13\,\dag}$   & $60^{\ddag}$          & $(23.4\pm0.5)\times10^{12\,\dag}$       & $60\pm2$      & $(7.2\pm0.1)\times10^{14}$    & $(16.4\pm0.1)\times10^{22}$ \\
G009.62$+$00.19 & $126\pm28$  & $(57.9\pm0.9)\times10^{14}$         & $114^{\ddag}$ & $(14.0\pm0.2)\times10^{13\,\dag}$     & $139\pm22$  & $(13.0\pm0.3)\times10^{13}$           & $149\pm3$   & $(10.3\pm0.1)\times10^{13}$                & $114\pm6$     & $(5.5\pm0.2)\times10^{14}$    & $(20.8\pm0.1)\times10^{22}$ \\
G010.47$+$00.02 & $140\pm2$   & $(9.0\pm0.1)\times10^{15}$          & $142^{\ddag}$ & $(6.5\pm0.2)\times10^{14\,\dag}$      & $109\pm2$   & $(8.3\pm0.4)\times10^{14}$            & $131\pm7$   & $(4.3\pm0.1)\times10^{14}$                 & $142\pm8$     & $(77.0\pm0.6)\times10^{14}$   & $(17.1\pm0.1)\times10^{22}$ \\
G010.62$-$00.38 & $56\pm5$    & $(18.0\pm0.3)\times10^{15}$         & $41^{\ddag}$  & $(11.0\pm0.2)\times10^{14\,\dag}$     & $41^{\ddag}$          & $(10.0\pm0.1)\times10^{14\,\dag}$  & $41^{\ddag}$          & $(9.1\pm0.1)\times10^{14\,\dag}$        & $41\pm3$      & $(7.3\pm0.2)\times10^{14}$    & $(40.9\pm0.1)\times10^{22}$ \\
G011.49$-$01.48 & $76\pm10$   & $(55.6\pm1.3)\times10^{14}$         & $35^{\ddag}$  & $2.3\times10^{14\,*}$                   & $35^{\ddag}$          & $1.3\times10^{14\,*}$                & $35^{\ddag}$          & $1.0\times10^{14\,*}$                     & $35\pm2$      & $(21.5\pm0.4)\times10^{13}$   & $(27.2\pm0.5)\times10^{21}$ \\
G011.91$-$00.61 & $142\pm15$  & $(29.6\pm0.3)\times10^{14}$         & $83^{\ddag}$  & $(3.6\pm0.3)\times10^{14\,\dag}$      & $83^{\ddag}$          & $(30.0\pm0.6)\times10^{13\,\dag}$  & $83^{\ddag}$          & $(12.4\pm0.2)\times10^{13\,\dag}$       & $83\pm3$      & $(62.3\pm0.7)\times10^{13}$   & $(55.5\pm0.4)\times10^{21}$ \\
G012.80$-$00.20 & $126^{\ddag}$         & $(11.0\pm0.3)\times10^{15\,\dag}$ & $126^{\ddag}$ & $(15.0\pm0.2)\times10^{14\,\dag}$     & $126^{\ddag}$         & $3.5\times10^{14\,*}$                & $126^{\ddag}$         & $(10.0\pm0.3)\times10^{14\,\dag}$       & $126\pm8$     & $(1.5\pm0.1)\times10^{14}$    & $(319.1\pm0.3)\times10^{21}$ \\
G012.88$+$00.48 & $137\pm13$  & $(13.0\pm0.7)\times10^{15}$         & $142^{\ddag}$ & $(9.9\pm0.2)\times10^{14\,\dag}$      & $158\pm26$  & $(1.5\pm0.1)\times10^{15}$            & $187\pm18$  & $(4.0\pm0.2)\times10^{15}$                 & $142\pm10$    & $(50.6\pm0.2)\times10^{13}$   & $(154.3\pm0.3)\times10^{21}$ \\
G012.90$-$00.24 & $38^{\ddag}$          & $(9.2\pm0.1)\times10^{14\,\dag}$ & $38^{\ddag}$  & $1.2\times10^{14\,*}$                   & $38^{\ddag}$          & $1.7\times10^{14\,*}$                & $38^{\ddag}$          & $1.1\times10^{14\,*}$                     & $38\pm1$      & $(27.2\pm0.2)\times10^{13}$   & $(169.2\pm0.5)\times10^{21}$ \\
G012.90$-$00.26 & $176\pm11$  & $(28.4\pm0.4)\times10^{15}$         & $129^{\ddag}$ & $(7.3\pm0.3)\times10^{14\,\dag}$      & $184\pm4$   & $(31.4\pm0.4)\times10^{14}$           & $194\pm11$  & $(4.5\pm0.4)\times10^{15}$                 & $129\pm6$     & $(4.0\pm0.1)\times10^{14}$    & $(57.5\pm0.4)\times10^{21}$ \\
G014.33$-$00.64 & $90\pm3$    & $(9.0\pm0.2)\times10^{14}$          & $71^{\ddag}$  & $(8.4\pm0.2)\times10^{13\,\dag}$      & $92\pm12$   & $(14.1\pm0.3)\times10^{13}$           & $74\pm7$    & $(28.9\pm0.3)\times10^{13}$                & $71\pm7$      & $(4.2\pm0.2)\times10^{14}$    & $(37.7\pm0.3)\times10^{21}$ \\
G015.03$-$00.67 & $127\pm19$  & $(5.8\pm0.5)\times10^{15}$          & $66^{\ddag}$  & $1.1\times10^{14\,*}$                   & $66^{\ddag}$          & $8.0\times10^{13\,*}$                & $66^{\ddag}$          & $4.3\times10^{13\,*}$                     & $66\pm4$      & $(51.8\pm0.2)\times10^{13}$   & $(116.3\pm0.3)\times10^{21}$ \\
G016.58$-$00.05 & $67\pm7$    & $(2.5\pm0.1)\times10^{14}$          & $59^{\ddag}$  & $(28.7\pm0.4)\times10^{13\,\dag}$     & $80\pm8$    & $(35.2\pm0.4)\times10^{13}$           & $59^{\ddag}$          & $(9.1\pm0.3)\times10^{13\,\dag}$        & $59\pm4$      & $(89.8\pm0.5)\times10^{13}$   & $(148.3\pm0.4)\times10^{21}$ \\
G023.00$-$00.41 & $87\pm2$    & $(18.1\pm0.4)\times10^{15}$         & $100^{\ddag}$ & $(2.2\pm0.2)\times10^{15\,\dag}$      & $104\pm5$   & $(2.4\pm0.2)\times10^{15}$            & $118\pm8$   & $(14.3\pm0.2)\times10^{14}$                & $101\pm6$     & $(34.7\pm0.7)\times10^{13}$   & $(188.1\pm0.6)\times10^{21}$ \\
G023.44$-$00.18 & $49\pm7$    & $(2.4\pm0.2)\times10^{14}$          & $49^{\ddag}$  & $(2.7\pm0.3)\times10^{14\,\dag}$      & $81\pm4$    & $(12.0\pm0.2)\times10^{13}$           & $49^{\ddag}$          & $(1.9\pm0.2)\times10^{14\,\dag}$        & $49\pm3$      & $(2.8\pm0.5)\times10^{14}$    & $(102.4\pm0.3)\times10^{21}$ \\
G027.36$-$00.16 & $118\pm8$   & $(35.3\pm0.4)\times10^{15}$         & $92^{\ddag}$  & $(1.7\pm0.1)\times10^{15\,\dag}$      & $104\pm6$   & $(10.5\pm0.3)\times10^{14}$           & $119\pm6$   & $(21.5\pm0.2)\times10^{14}$                & $92\pm10$     & $(9.0\pm0.4)\times10^{14}$    & $(119.2\pm0.3)\times10^{21}$ \\
G028.86$+$00.06 & $151\pm6$   & $(17.0\pm0.3)\times10^{14}$         & $68^{\ddag}$  & $(16.4\pm0.3)\times10^{13\,\dag}$     & $69^{\ddag}$          & $5.2\times10^{13\,*}$                & $69^{\ddag}$          & $(11.0\pm0.5)\times10^{12\,\dag}$       & $69\pm6$      & $(91.0\pm0.6)\times10^{13}$   & $(153.1\pm0.4)\times10^{21}$ \\
G029.95$-$00.01 & $119\pm27$  & $(13.7\pm0.2)\times10^{14}$         & $135^{\ddag}$ & $(1.8\pm0.1)\times10^{14\,\dag}$      & $147\pm12$  & $(20.2\pm0.4)\times10^{13}$           & $158\pm2$   & $(13.3\pm0.2)\times10^{13}$                & $135\pm11$    & $(53.6\pm0.3)\times10^{13}$   & $(116.2\pm0.4)\times10^{21}$ \\
G031.28$+$00.06 & $70\pm9$    & $(16.5\pm0.4)\times10^{14}$         & $36^{\ddag}$  & $(16.2\pm0.2)\times10^{13\,\dag}$     & $41\pm6$    & $(10.1\pm0.2)\times10^{13}$           & $53\pm6$    & $(8.3\pm0.2)\times10^{13}$                 & $36\pm2$      & $(2.1\pm0.4)\times10^{14}$    & $(103.1\pm0.4)\times10^{21}$ \\
G031.58$+$00.07 & $61\pm3$    & $(13.2\pm0.2)\times10^{15}$         & $42^{\ddag}$  & $(2.6\pm0.1)\times10^{15\,\dag}$      & $42^{\ddag}$          & $(3.3\pm0.1)\times10^{15\,\dag}$   & $42^{\ddag}$          & $(3.7\pm0.1)\times10^{15\,\dag}$        & $42\pm5$      & $(74.7\pm0.7)\times10^{13}$   & $(96.5\pm0.3)\times10^{21}$ \\
G032.04$+$00.05 & $77\pm1$    & $(50.4\pm0.1)\times10^{14}$         & $53^{\ddag}$  & $(5.7\pm0.1)\times10^{14\,\dag}$      & $91\pm3$    & $(2.3\pm0.2)\times10^{14}$            & $88\pm6$    & $(13.4\pm0.2)\times10^{13}$                & $53\pm3$      & $(27.8\pm0.5)\times10^{13}$   & $(70.8\pm0.2)\times10^{21}$ \\
G034.39$+$00.22 & $39^{\ddag}$          & $(10.8\pm0.7)\times10^{13\,\dag}$ & $39^{\ddag}$  & $(22.2\pm0.3)\times10^{12\,\dag}$     & $58\pm2$    & $(10.5\pm0.1)\times10^{12}$           & $39^{\ddag}$          & $(10.4\pm0.3)\times10^{12\,\dag}$       & $39\pm4$      & $(5.8\pm0.3)\times10^{13}$    & $(69.6\pm0.3)\times10^{21}$ \\
G035.02$+$00.34 & $64\pm8$    & $(2.0\pm0.1)\times10^{15}$          & $40^{\ddag}$  & $(3.0\pm0.1)\times10^{14\,\dag}$      & $40^{\ddag}$          & $6.9\times10^{13\,*}$                & $40^{\ddag}$          & $4.2\times10^{13\,*}$                     & $40\pm4$      & $(7.5\pm0.4)\times10^{13}$    & $(85.9\pm0.5)\times10^{21}$ \\
G035.19$-$00.74 & $58\pm3$    & $(11.6\pm0.2)\times10^{14}$         & $66^{\ddag}$  & $(20.4\pm0.2)\times10^{13\,\dag}$     & $86\pm6$    & $(10.8\pm0.4)\times10^{13}$           & $78\pm4$    & $(10.3\pm0.6)\times10^{13}$                & $66\pm4$      & $(40.9\pm0.8)\times10^{13}$   & $(45.8\pm0.2)\times10^{21}$ \\
G035.20$-$01.73 & $85\pm4$    & $(1.6\pm0.1)\times10^{15}$          & $47^{\ddag}$  & $(10.5\pm0.2)\times10^{13\,\dag}$     & $47^{\ddag}$          & $5.1\times10^{13\,*}$                & $47^{\ddag}$          & $2.2\times10^{13\,*}$                     & $47\pm3$      & $(6.9\pm0.1)\times10^{13}$    & $(62.8\pm0.2)\times10^{21}$ \\
G037.43$+$01.51 & $44\pm5$    & $(4.2\pm0.2)\times10^{15}$          & $44^{\ddag}$  & $(4.5\pm0.1)\times10^{14\,\dag}$      & $80\pm5$    & $(27.2\pm0.3)\times10^{13}$           & $76\pm3$    & $(23.3\pm0.3)\times10^{13}$                & $43\pm3$      & $(6.7\pm0.1)\times10^{13}$    & $(248.3\pm0.4)\times10^{21}$ \\
G043.16$+$00.01 & $106\pm10$  & $(6.1\pm0.4)\times10^{15}$          & $72^{\ddag}$  & $(5.3\pm0.3)\times10^{14\,\dag}$      & $72^{\ddag}$          & $(21.2\pm0.3)\times10^{13\,\dag}$  & $72^{\ddag}$          & $(21.5\pm0.3)\times10^{13\,\dag}$       & $72\pm6$      & $(4.0\pm0.1)\times10^{14}$    & $(97.4\pm0.4)\times10^{21}$ \\
G043.79$-$00.12 & $53\pm5$    & $(4.3\pm0.1)\times10^{15}$          & $51^{\ddag}$  & $(4.8\pm0.1)\times10^{14\,\dag}$      & $54\pm2$    & $(18.6\pm0.2)\times10^{13}$           & $51^{\ddag}$          & $(10.6\pm0.1)\times10^{13\,\dag}$       & $51\pm3$      & $(84.5\pm0.3)\times10^{13}$   & $(159.2\pm0.8)\times10^{21}$ \\
G049.48$-$00.36 & $104\pm15$  & $(2.9\pm0.1)\times10^{15}$          & $124^{\ddag}$ & $(29.3\pm0.5)\times10^{13\,\dag}$     & $112\pm3$   & $(10.9\pm0.4)\times10^{13}$           & $134\pm11$  & $(11.9\pm0.3)\times10^{13}$                & $124\pm7$     & $(21.9\pm0.4)\times10^{14}$   & $(30.1\pm0.1)\times10^{22}$ \\
G049.48$-$00.38 & $134\pm20$  & $(9.9\pm0.1)\times10^{14}$          & $134^{\ddag}$ & $(8.4\pm0.3)\times10^{13\,\dag}$      & $141\pm4$   & $(2.2\pm0.2)\times10^{14}$            & $155\pm3$   & $(10.1\pm1.0)\times10^{13}$                & $134\pm5$     & $(55.4\pm0.5)\times10^{14}$   & $(32.5\pm0.2)\times10^{21}$ \\
G059.78$+$00.06 & $70\pm4$    & $(10.8\pm0.4)\times10^{14}$         & $39^{\ddag}$  & $(21.2\pm0.2)\times10^{12\,\dag}$     & $39^{\ddag}$          & $1.7\times10^{13\,*}$                & $39^{\ddag}$          & $1.1\times10^{13\,*}$                     & $39\pm1$      & $(44.5\pm0.5)\times10^{13}$   & $(78.3\pm0.2)\times10^{21}$ \\
G069.54$-$00.97 & $79\pm7$    & $(2.8\pm0.2)\times10^{15}$          & $65^{\ddag}$  & $(23.2\pm0.4)\times10^{13\,\dag}$     & $44\pm5$    & $(10.7\pm0.2)\times10^{13}$           & $65^{\ddag}$          & $(9.0\pm0.2)\times10^{13\,\dag}$        & $65\pm4$      & $(75.0\pm0.3)\times10^{13}$   & $(39.7\pm0.3)\times10^{21}$ \\
G075.76$+$00.33 & $40^{\ddag}$         & $(3.2\pm0.4)\times10^{14\,\dag}$                        & $40^{\ddag}$ & $(1.2\pm0.1)\times10^{14\,\dag}$                             & $40^{\ddag}$         & $8.1\times10^{13\,*}$                           & $40^{\ddag}$         & $4.3\times10^{13\,*}$                                & $40\pm3$           & $(9.7\pm0.3)\times10^{14}$                  & $(27.4\pm0.2)\times10^{21}$ \\
G078.12$+$03.63 & $139\pm4$   & $(35.9\pm0.4)\times10^{14}$         & $113^{\ddag}$ & $4.0\times10^{14\,*}$                   & $113^{\ddag}$         & $1.1\times10^{14\,*}$                & $113^{\ddag}$         & $8.6\times10^{13\,*}$                     & $113\pm13$    & $(3.4\pm0.1)\times10^{14}$    & $(74.8\pm0.2)\times10^{21}$ \\
G081.75$+$00.59 & $44\pm2$    & $(4.1\pm0.4)\times10^{14}$          & $44^{\ddag}$  & $(21.1\pm0.5)\times10^{12\,\dag}$     & $44^{\ddag}$          & $8.3\times10^{13\,*}$                & $44^{\ddag}$          & $7.7\times10^{13\,*}$                     & $44\pm2$      & $(73.4\pm1.0)\times10^{13}$   & $(114.2\pm0.4)\times10^{21}$ \\
G081.87$+$00.78 & $136\pm7$   & $(21.8\pm0.4)\times10^{14}$         & $109^{\ddag}$ & $(18.8\pm0.4)\times10^{13\,\dag}$     & $127\pm6$   & $(10.3\pm0.3)\times10^{13}$           & $137\pm9$   & $(10.6\pm0.4)\times10^{13}$                & $109\pm9$     & $(31.0\pm0.3)\times10^{13}$   & $(28.9\pm0.2)\times10^{21}$ \\
G092.67$+$03.07 & $53\pm3$    & $(14.7\pm0.2)\times10^{14}$         & $54^{\ddag}$  & $(14.2\pm0.4)\times10^{13\,\dag}$     & $54^{\ddag}$          & $8.7\times10^{13\,*}$                & $54^{\ddag}$          & $(7.2\pm0.7)\times10^{13\,\dag}$        & $54\pm3$      & $(75.0\pm0.3)\times10^{13}$   & $(103.2\pm0.2)\times10^{21}$ \\
G109.87$+$02.11 & $78\pm5$    & $(1.7\pm0.1)\times10^{14}$          & $63^{\ddag}$  & $(11.5\pm0.3)\times10^{12\,\dag}$     & $63^{\ddag}$          & $1.0\times10^{13\,*}$                & $63^{\ddag}$          & $1.1\times10^{13\,*}$                     & $63\pm5$      & $(65.2\pm0.5)\times10^{13}$   & $(64.5\pm0.2)\times10^{21}$ \\
G111.54$+$00.77 & $107\pm4$   & $(8.9\pm0.2)\times10^{15}$          & $80^{\ddag}$  & $(8.3\pm0.2)\times10^{14\,\dag}$      & $84\pm8$    & $(24.4\pm0.3)\times10^{13}$           & $97\pm6$    & $(10.7\pm0.1)\times10^{13}$                & $80\pm7$      & $(61.2\pm0.6)\times10^{13}$   & $(38.0\pm0.1)\times10^{21}$ \\
G121.29$+$00.65 & $49\pm5$    & $(8.4\pm0.4)\times10^{14}$          & $32^{\ddag}$  & $(22.8\pm0.3)\times10^{12\,\dag}$     & $69\pm4$    & $(1.9\pm0.2)\times10^{13}$            & $32^{\ddag}$          & $1.6\times10^{13\,*}$                     & $32\pm2$      & $(57.1\pm1.0)\times10^{13}$   & $(31.6\pm0.2)\times10^{21}$ \\
G123.06$-$06.30 & $63\pm10$   & $(2.4\pm0.1)\times10^{15}$          & $42^{\ddag}$  & $(32.9\pm0.5)\times10^{13\,\dag}$     & $70\pm2$    & $(17.4\pm0.2)\times10^{13}$           & $42^{\ddag}$          & $(10.0\pm0.1)\times10^{13\,\dag}$       & $42\pm2$      & $(94.4\pm0.2)\times10^{13}$   & $(81.2\pm0.3)\times10^{21}$ \\
G133.94$+$01.06 & $137^{\ddag}$         & $(11.7\pm0.4)\times10^{14\,\dag}$ & $129^{\ddag}$ & $(13.4\pm0.4)\times10^{14\,\dag}$     & $144\pm4$   & $(10.1\pm0.1)\times10^{14}$           & $154\pm3$   & $(10.4\pm0.2)\times10^{14}$                & $129\pm10$    & $(11.5\pm0.3)\times10^{13}$   & $(14.5\pm0.3)\times10^{21}$ \\
G168.06$+$00.82 & $-$         & $-$                        & $-$ & $-$                             & $-$         & $-$                          & $-$         & $-$                               & $-$           & $-$                  & $(24.4\pm0.1)\times10^{21}$ \\
G176.51$+$00.20 & $34\pm3$    & $(2.5\pm0.1)\times10^{14}$          & $30^{\ddag}$  & $(16.3\pm0.2)\times10^{12\,\dag}$     & $30^{\ddag}$          & $1.2\times10^{13\,*}$                & $30^{\ddag}$          & $(1.7\pm0.2)\times10^{13\,\dag}$        & $30\pm1$      & $(89.9\pm0.1)\times10^{13}$   & $(11.3\pm0.1)\times10^{21}$ \\
G183.72$-$03.66 & $84\pm6$    & $(23.5\pm0.4)\times10^{13}$         & $54^{\ddag}$  & $(3.7\pm0.3)\times10^{13\,\dag}$      & $54^{\ddag}$          & $5.1\times10^{13\,*}$                & $96\pm4$    & $(5.2\pm0.2)\times10^{13}$                 & $54\pm3$      & $(13.8\pm0.5)\times10^{13}$   & $(25.3\pm0.2)\times10^{21}$ \\
G188.94$+$00.88 & $51\pm9$    & $(12.6\pm0.2)\times10^{13}$         & $35^{\ddag}$  & $(1.4\pm0.1)\times10^{13\,\dag}$      & $35^{\ddag}$          & $1.1\times10^{13\,*}$                & $35^{\ddag}$          & $1.1\times10^{13\,*}$                     & $35\pm2$      & $(27.8\pm0.8)\times10^{13}$   & $(41.8\pm0.2)\times10^{21}$ \\
G192.60$-$00.04 & $96\pm7$    & $(10.7\pm0.5)\times10^{15}$         & $80^{\ddag}$  & $(10.7\pm0.1)\times10^{14\,\dag}$     & $130\pm14$  & $(21.3\pm0.3)\times10^{13}$           & $80^{\ddag}$          & $1.3\times10^{14\,*}$                     & $80\pm8$      & $(43.4\pm0.5)\times10^{13}$   & $(17.6\pm0.1)\times10^{21}$ \\
G209.00$-$19.38 & $106\pm9$   & $(1.3\pm0.1)\times10^{14}$          & $126^{\ddag}$ & $(1.9\pm0.1)\times10^{14\,\dag}$      & $123\pm7$   & $(11.2\pm0.3)\times10^{13}$           & $133\pm6$   & $(10.5\pm0.5)\times10^{13}$                & $126\pm15$    & $(2.4\pm0.1)\times10^{13}$    & $(27.5\pm0.2)\times10^{21}$ \\
G232.62$+$00.99 & $51\pm4$    & $(10.2\pm0.1)\times10^{13}$         & $34^{\ddag}$  & $2.0\times10^{13\,*}$                   & $34^{\ddag}$          & $1.4\times10^{13\,*}$                & $34^{\ddag}$          & $1.2\times10^{13\,*}$                     & $34\pm2$      & $(17.4\pm0.4)\times10^{13}$   & $(20.6\pm0.1)\times10^{21}$ \\
\bottomrule
\end{tabular}

\begin{tablenotes}[flushleft]
\scriptsize
\item[$-$] indicates that neither $T_{\rm rot}$ nor $N_{\rm tot}$ could be constrained for the corresponding species.
\item[$^{a}$] H$_2$ column densities are adopted from \citet{Luo2024}. They were derived from C$^{18}$O $J= 1$--0 observed in the same IRAM-30\,m data set, using $N({\rm H_2}) = 4.37 \times 10^6 N({\rm C^{18}O})$~\citep{Frerking1982}. Since C$^{18}$O $J=1$--0 lies at 109.782\,GHz, these values correspond to a beam size of about 22.4$''$ and provide a homogeneous beam-averaged H$_2$ column-density scale for source-to-source comparison.
\item[$^{*}$] Upper limits derived for non-detections.
\item[$^{\dag}$] $N_{\rm tot}$ computed by adopting a fixed $T_{\rm rot}$ because fewer than three usable transitions were available for an independent rotation-diagram fit.
\item[$^{\ddag}$] $T_{\rm rot}$ adopted from CH$_3$CN as a proxy excitation temperature.
\end{tablenotes}
\end{threeparttable}
\end{sidewaystable*}

\end{appendix} 
\end{document}